\documentclass[%
reprint,
 amsmath,amssymb,
 aps,
pra,
]{revtex4-2}

\usepackage{CJKutf8}
\usepackage{graphicx}
\usepackage{dcolumn}
\usepackage{bm}

\usepackage{xcolor}

\begin{document}

\begin{CJK}{UTF8}{gbsn}

\title{Higher-order valley vortices enabled by synchronized rotation in a photonic crystal} 
 
\author{{Rui Zhou}\textsuperscript{1}}

\author{{Hai Lin}\textsuperscript{1}}

\email{Corresponding author: linhai@mail.ccnu.edu.cn}

\author{{Yanjie Wu}\textsuperscript{1}}

\author{{Zhifeng Li}\textsuperscript{1}}

\author{{Zihao Yu}\textsuperscript{1}}

\author{{Y. Liu(刘泱杰)}\textsuperscript{2,3}}
\email{Corresponding author: yangjie@hubu.edu.cn}

\author{{Dong-Hui Xu}\textsuperscript{2}}
 
\affiliation{$^{1}$ College of Physics Science and Technology, Central China Normal University, Wuhan 430079, Hubei Province\\
$^{2}$School of Physics and Electronic Sciences, Hubei University, Wuhan 430062, Hubei Province\\
$^{3}$Lanzhou Center for Theoretical Physics, Key Laboratory of Theoretical Physics of Gansu Province, Lanzhou University, Lanzhou 730000, Gansu Province}

\date{To be submitted to~\emph{Photon. Res.}, \today, v23}
             
\begin{abstract}
Synchronized rotation of unit cells in a periodic structure provides a novel design perspective for manipulation of band topology. We then design a two-dimensional version of higher-order topological insulators (HOTI), by such rotation in a triangular photonic lattice with $\mathcal{C}_3$ symmetry. This HOTI supports the hallmark zero-dimensional corner states and simultaneously the one-dimensional edge states. We also find that our photonic corner states carry chiral orbital angular momenta locked by valleys, whose wavefunctions are featured by the phase vortex (singularity) positioned at the maximal Wyckoff points. Moreover, when excited by a fired source with various frequencies, the valley topological states of both one-dimensional edges and zero-dimensional corners emerge simultaneously. Extendable to higher or synthetic dimensions, our work provides access to a chiral vortex platform for HOTI realisations in the THz photonic system.  
 
\end{abstract}

\maketitle
\end{CJK}

\section{\label{sec:level1}
Introduction}

With further developments of photonic crystals (PhC) armored by topological understanding for condensed matters,  people have found miscellaneous photonic counterparts of topological phases~\cite{2008Possible,2008Reflection,2013Photonic,2017Valley,2015}. The topological edge states generated on the interface between different topological phases promise superior features such as robustly smooth transmission, backscattering suppression and defect immunity despite rather strong perturbation of the local boundary. Following quantum Hall (QH) phases~\cite{2008Reflection}, more intricate topological phases such as quantum spin Hall (QSH) phases~\cite{2013Photonic, 2015}, and quantum valley Hall (QVH) phases~\cite{2017Valley} are also invented in the context of analogue PhC systems, the two of which respectively exploit the dichroism freedom by pseudo-spin/valley concepts in classical wave setups. Such configurable symmetrical lattices furthermore provide easy access to topological crystalline insulators (TCI), for example those with synchronous rotation giving rise to high tunabilility in practical realisation~\cite{ZhouR2021}. Specifically for a $\mathcal{C}_{3}$ Kagome lattice of broken inversion symmetry, distinct valley states will emerge in the first Brillouin zone (FBZ), and produce their Berry curvature of opposite values~\cite{2017Valley,2018Valleychen,2018Valleyzhang}. Such valleytronics concept calls for bulk valley states locked to their chiralities, which are possible to couple into and out of communication devices such as valley filters and valley sources respectively~\cite{2017Valley,2018Valleychen,2018Valleyzhang,2017Observation,2017Valleydong}. 

Nevertheless, a concept of corner states from higher-order topology that is one further dimension lower than the edges in a two-dimensional (2D) setup~\cite{Benalcazar2017a,2017Electric,2017d,Schindler2018, 2019Quantization}, has added new bricks to the premise for topological information devices. Among the class of higher-order phase, one type of topology is measured by the fractional bulk polarization (or the position of Wannier centers)~\cite{2019Quantization}. For instance, zero-dimensional (0D) corner states, other than the one-dimensional (1D) edge ones, will emerge in the second-order TCIs, whose spacial positions are associated with Wannier centers determined from the polarization value~\cite{0Nonlinear,2018Second,2020Multipolar,2021Dual}.

Peculiar to the classical analogue for topological quantum physics, the spatial vortex, i.e. the wavefunction with undefined phase in certain spatial positions, remains yet less explored despite its mechanical power to manipulate macroparticles. The vortex flow of electromagnetic waves, also defined as the orbital angular momenta (OAM) of light,
may open up new avenues to exert optical torques to matters in a non-invasive manner. Such a possibility shall be revealed in this paper, where we will design and demonstrate a valley higher-order topological insulator (HOTI) in a triangular lattice with $\mathcal{C}_{3}$ symmetry, fueled by synchronous rotation of each unit cell. By observing the phase of electric fields near $K$ and $K^{\prime}$ points we recognise a valley selection feature discussed previously~\cite{2017Valley}. It is also found that the synchronous rotation mechanism of unit cells induces a band inversion at valleys, which leads to a topological phase transition in our photonic system. This topological transition can be characterized by the extended 2D bulk polarization related to Zak phase~\cite{2017Novel,2018zero,2019Quantization}. In the electric field of the valley HOTI, pointwise corner states are predicted by the 2D bulk polarization. Furthermore, not only does our proposed HOTI have a vortex edge state locked to one of dichroic valleys~\cite{HeX2019,MaoY2021}, but also supports a topologically corner states. Using chiral point sources of different frequencies, our simulations verify that the electromagnetic waves shape into high-quality corner states and robust edge states. Our idea can be extended to higher or synthetic dimensions, which contributes to an experimentally feasible platform for HOTI in the photonic vortex system~\cite{2018Valleyzhang,2017Valley,2017Observation,2019All,2020Higher,2021Valley}.

\section{\label{sec:level2}Theory and model}
 
We propose a 2D PhC in triangular lattice with $\mathcal{C}_{3{v}}$ symmetry, the unit cell of which is composed of six identical pure dielectric cylinders embedded in air, as shown in the left panel of Fig.~1(a). And the maximal Wyckoff points in the unit cell is represented by labels $\mathrm{o}$, $\mathrm{p}$, $\mathrm{q}$ in real space. The dielectric permittivity is $\varepsilon_{d}=7.5$, $a_{0}$ is the lattice constant, and $\mathbf{a}_{1}$ and $\mathbf{a}_{2}$ are the lattice vectors, with cylinders diameter $d=0.2a_{0}$, the lattice constant $a_{0} =50{\rm \mu m}$, and $a_{0}/R=3.5$. The synchronous rotation angle of the dielectric cylinders in the unit cell is represented by $\theta$, shown in the right panel of Fig.~1(a), with counterclockwise rotation as the positive direction of rotation whose maximum rotation angle is $60^{\circ}$. 

\begin{figure}
\includegraphics[width=0.47\textwidth]{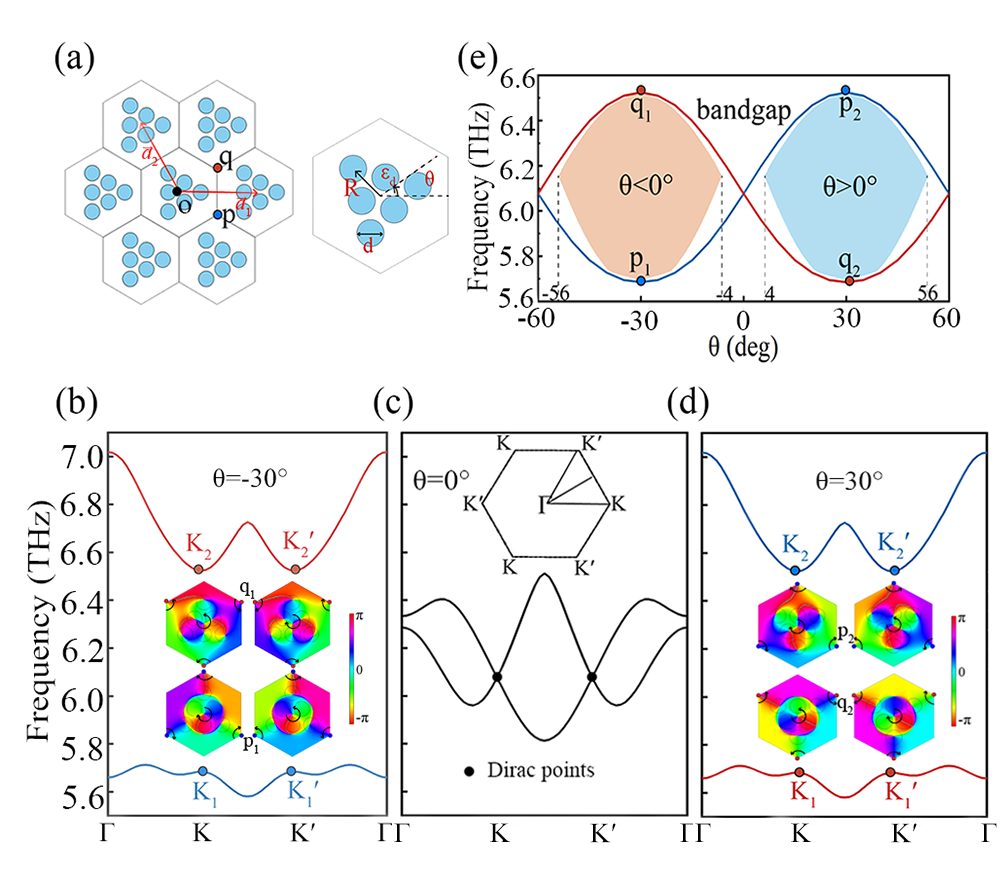}
\caption{\label{fig:epsart}(a) Left: Schematic of un-rotated sampled PhC with lattice constant $a_{0}$ where the three positions in the $\mathcal{C}_{3}$ point group is labelled by $\mathrm{o}$, $\mathrm{p}$, $\mathrm{q}$ respectively. Right: rotated unit cell with $\theta$ as the rotation angle. (b-d) Dispersion bands of the valley PC with $\theta=-30^{\circ},0^{\circ}$ and $30^{\circ}$ [insets of (b, d) shows the phase distributions]. Valley points of lower and higher frequency are labelled by ${K}_{1}\left({K}_{1}^{\prime}\right)$ and ${K}_{2}\left({K}_{2}^{\prime}\right)$ respectively. (b) When $\theta=-30^{\circ}$, at the $\mathrm{p}_{1}$ and $\mathrm{q}_{1}$ points the phase distributions reveal that ${K}_{1}\left(_{1}^{\prime}\right)$ and ${K}_{2}\left({K}_{2}^{\prime}\right)$ have opposite chirality, whose handedness is indicated by the black arrow in insets. (c) When $\theta=0^{\circ}$,  the Dirac points appear at points ${K}$ and ${K}^{\prime}$ in the FBZ, and the inset shows the FBZ of the triangular lattice. (d) When $\theta=30^{\circ}$, ${K}$ and ${K}^{\prime}$ valley points near the band gap is reversed in frequency order at $\mathrm{q}_{2}$ and $\mathrm{p}_{2}$ compared to panel(b). (e) The blue and the red band represent the frequency gap variation of ${K}$-valley when the unit cell rotates between $-60$ to $+60^\circ$ in a period. The crayon and orange shading indicate the complete band gap width of bands when rotating for different angles. Note that no gaps remain between $4^\circ$.}
\end{figure}

In this work, a finite element method (FEM) is used to calculate the PhC dispersion and to solve for the related electric fields. In a $\mathcal{C}_3$-symmetric lattice, the photonic FBZ contains a pair of $K$ and $K^{\prime}$ points in its vertices, which is named as valley points~\cite{2009electronic,2016Valleytronics}, as shown in Fig.~1(c) inset. Here the valley states at ${K}$ and ${K}^{\prime}$, connected by TR symmetry~\cite{2014photonics, 2018VD}, are both linearly-dispersed, which are hence named as Dirac points~\cite{DC2021}. We shall only focus on eigenstates near valley points, and refer valleys ${K}$ and ${K}^{\prime}$ to \emph{Dirac points}, throughout our whole paper to be succinct. Considering the transverse magnetic (TM) mode for simplicity, the band degeneracy at two Dirac points in Fig.~1(c) is levitated away from linear dispersion, by rotating the dielectric cylinders in every unit, which are shown in panels (b, d). For complete band diagram see Sec.~$\mathrm{I}$ of Supplemental Material~\cite{SM}. To be specific, when rotated away from the original lattice $\left(\theta=0^{\circ}\right)$ in panel (c), the Dirac degeneracy is levitated to open a bandgap near the Dirac points. We define the lower and higher frequency state at ${K}$(${K}^{\prime}$) as represented by ${K}_{1}\left({K}_{1}^{\prime}\right)$ and ${K}_{2}\left({K}_{2}^{\prime}\right)$, respectively in panels (b, d). When the unit cells are rotated clockwise $\theta=-30^{\circ}$, two pairs of valley states are presented as insets of Fig.~1(d). And these valley states in gap occupy chirality in the sense of
circular-polarized OAM, which is manifest by the phase
distribution of $\mathrm{E}_{z}$, i.e., $\arg({E}_{z})$~\cite{2017ValleyD,2017Valleydong}. For ${K}$ valley, the phases of ${K}_{1}$ and ${K}_{2}$ have opposite vortex chirality at the positions of $\mathrm{p}$ and $\mathrm{q}$ respectively [denoted as $\mathrm{p}_{1}$ and $\mathrm{q}_{1}$ for $\theta=+30^{\circ}$ and $\mathrm{p}_{2}$ and $\mathrm{q}_{2}$ for $\theta=-30^{\circ}$ shown in insets of Fig.~1(b, d)], and vice versa for ${K}^{\prime}$ valley. With the opposite signs of $\theta$, the frequency order of the valleys corresponding to $\mathrm{p}$ and $\mathrm{q}$ positions are reverse as shown in Fig.~1(e), indicating a typical band inversion that leads to a topological phase transition~\cite{2015}. The crayon and orange shadings in panel (e) mark out the bandgap width of the system during synchronous rotation of unit cells. 

\begin{figure}
\includegraphics[width=0.5\textwidth]{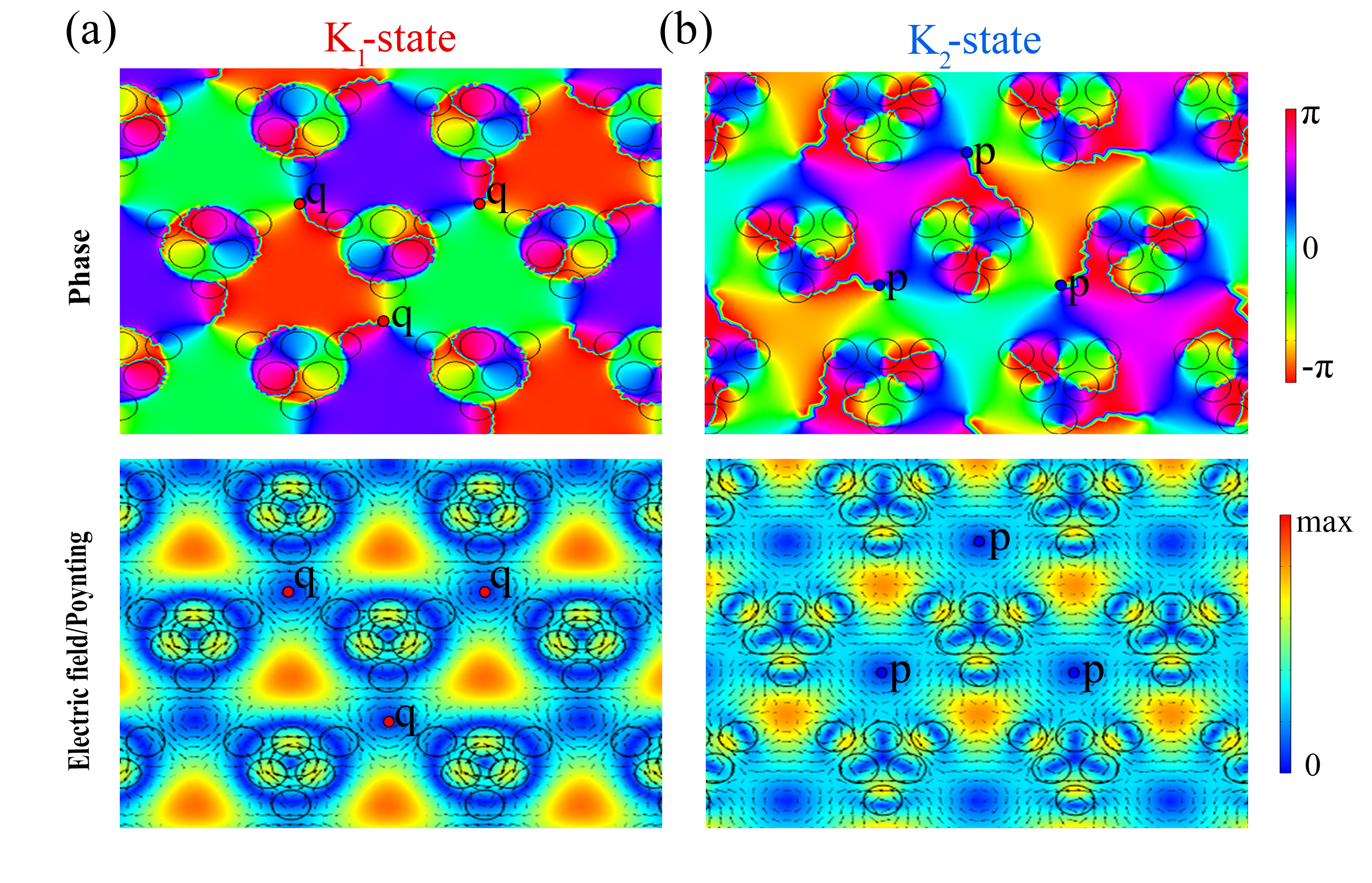}
\caption{\label{fig:epsart} Electric field distribution $\vert {E}_{z}\vert (x, y)$ of the $K$-valley state (low frequency ${K}_{1}$, high frequency ${K}_{2}$) at positions p, q (p, q indicate positions with $\mathcal{C}_{3{v}}$ symmetry). Parameter: rotational angle $\theta=30^{\circ}$. The upper and lower panels respectively represent the valley phase and electric field amplitude distribution of the large-period lattice, the arrows in the lower panel indicate the corresponding time-averaged Poynting vector. } 
\end{figure}

Let us focus on the properties of the $K$-valley state, i.e. ${K}_{1}$ and ${K}_{2}$ for its lower and higher band in frequency respectively, while the counterparts for $K^{\prime}$ valley can be deduced by TR symmetry (\emph{cf}. Sec.~$\mathrm{II}$ of Supplemental Material~\cite{SM})~\cite{2014photonics,2018Second,Higher}. We find that the photonic valley states are chiral in the sense of phase singularity, which can be readily seen
from the electric fields in Fig.~2, where the top and bottom
panels display the phase and amplitude distributions, respectively. In the positions of maximal Wyckoff~\cite{2017d}, q and p, the electric amplitudes $E_z$ vanish and thus the phases become singular
for the chiral valley states~\cite{Jan2016Pseudospin}. Note that in our PhC unit of $\mathcal{C}_3$ symmetry, it has three maximal Wyckoff positions: $\mathrm{o}$ at the center of the unit cell, and $\mathrm{q}$ and $\mathrm{p}$ at the vertices of it~(\emph{cf}.~Fig.~1(a) and Sec.~$\mathrm{IV}$ of Supplemental Material~\cite{SM}). The electric fields above reveal a typical feature of vortex field, aligning in flow directions defined by time-averaged Poynting vectors $\mathbf{S}=\operatorname{Re}\left[\mathbf{E} \times \mathbf{H}^{*}\right] / 2$~\cite{Litvin2011Poynting} shown
in the arrows of the lower panels of Fig.~2. Therefore we can control the chirality of the valley vortex by choosing the source chirality. Other than such valley-chirality locking, we also note that $K_1$-state in $E_z$ field distribution in panel (a) actually supports a whole circle of zero amplitude and singularity, and that in panel (b) a Y-type singularity curve, other than discrete singularity points.  

Recently, it has been suggested that the HOTI state can be evaluated by integrating Berry connection in the FBZ, which is actually the Zak phase along the wave vector direction~\cite{2017Novel,2018zero,2018Second,Higher}. The 2D Zak phase is connected to the fractional polarization through $\theta_{i}=2\pi P_{i}$ for $i=1, 2$, where its Zak phase or polarization is completely determined by the bulk property. 
In the 2D system, the bulk polarization is defined in terms of Berry connection as~\cite{2021H-O}:
\begin{equation}\label{Pi}
P_{i}=-\frac{1}{(2 \pi)^{2}} \int d^{2} \boldsymbol{k} \operatorname{Tr}\left[\hat{\mathcal{A}}_{i}\right], \quad i=1, 2
\end{equation}
with $i$ indicating the component of $\mathbf{P}$ along the reciprocal lattice vector $\mathbf{b}_{i}(i=1,2)$. Here $\left[A_{i}(\mathbf{k})\right]^{m n}=-i\left\langle u^{m}(\mathbf{k})\left|\partial k_{i}\right| u^{n}(\mathbf{k})\right\rangle$ is the Berry connection matrix where $m$ and $n$ run over occupied energy bands, and  $\left|u^{n}(\mathbf{k})\right\rangle$ is the periodic Bloch function for the $n$-th band with $\mathbf{k}=k_{1} \mathbf{b}_{1}+k_{2} \mathbf{b}_{2}$ as the wave vector, where $k_1, k_2$ are integers. In reciprocal space, we can express the polarization in terms of lattice vector via a numerical integration in Eq.~\ref{Pi}. Then the bulk polarization $\mathbf{P}$ of our PhC along $\mathbf{b}_{1,2}$, as illustrated in Fig.~3(a), equaling $\left({1}/{3}, {1}/{3}\right)$ or $\left(-{1}/{3},-{1}/{3}\right)$ for $\theta \in\left(10^{\circ}, 30^{\circ}\right)$ and $\theta \in\left(-10^{\circ}, -30^{\circ}\right)$, indicates the topologically nontrivial phase, while $\left(0, 0\right)$ for $\theta \in\left(-10^{\circ}, 10^{\circ}\right)$ a trivial phase (\emph{cf}.~Sec.~$\mathrm{III}$ of Supplemental Material~\cite{SM}). Therein the blue solid circle $\mathrm{P}_{1}$ and the red hollow circle $\mathrm{P}_{2}$ represent the polarization values in $\mathbf{b}_{1}$ and $\mathbf{b}_{2}$ directions. When $\theta \in\left(-10^{\circ}, -30^{\circ}\right)$, $\mathbf{P}=\left(-{1}/{3},-{1}/{3}\right)$ this means that the Wannier center is located at the maximal Wyckoff position p as shown in Fig.~3(b). When $\theta \in\left(10^{\circ}, 30^{\circ}\right)$, $\mathbf{P}=\left({1}/{3}, {1}/{3}\right)$, Wannier center is pinned to the maximal Wyckoff position q as shown in Fig.~3(c) inset. As the pillars in unit cells rotate, the topologically nontrivial polarization $\mathbf{P}$ varies from $\left(-{1}/{3},-{1}/{3}\right)$ to $\left({1}/{3}, {1}/{3}\right)$. Consequently, the corner states associated with different polarization values appear at the maximal Wyckoff positions due to valley selection.

\begin{figure}
\includegraphics[width=0.5\textwidth]{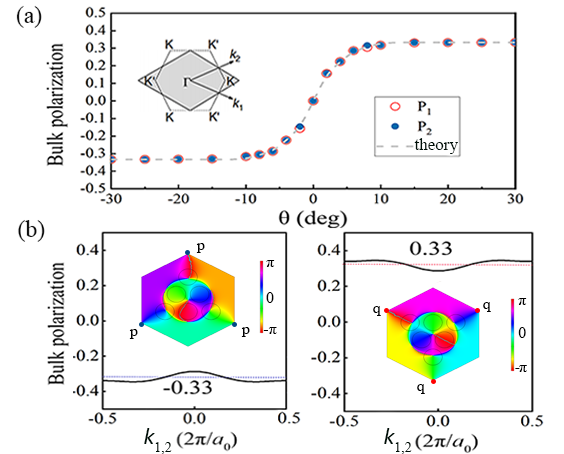} 
\caption{\label{fig:epsart}(a) Bulk polarization changes when the unit cells rotate synchronously. Red circles for $\mathrm{P}_{1}$, blue dots for $\mathrm{P}_{2}$, dotted line for the theoretical calculation, and inset for schematic of the FBZ. (b) Left: when $\theta=-30^{\circ}$, the polarization value along the wave vector $\mathbf{k}_{2}$ direction $\mathrm{P}_{2}={1}/{3}$ [bulk polarization $\mathbf{P}=\left(-{1}/{3},-{1}/{3}\right)$] where Wannier center in unit cell aligns at the maximal Wyckoff positions $p$ (blue dots of inset). Right: when $\theta=30^{\circ}$, bulk polarization $\mathbf{P}=\left({1}/{3}, {1}/{3}\right)$ where Wannier center is located at the maximal Wyckoff position $q$ (red dots of inset).}
\end{figure}

\begin{figure*}
\includegraphics[width=1\textwidth]{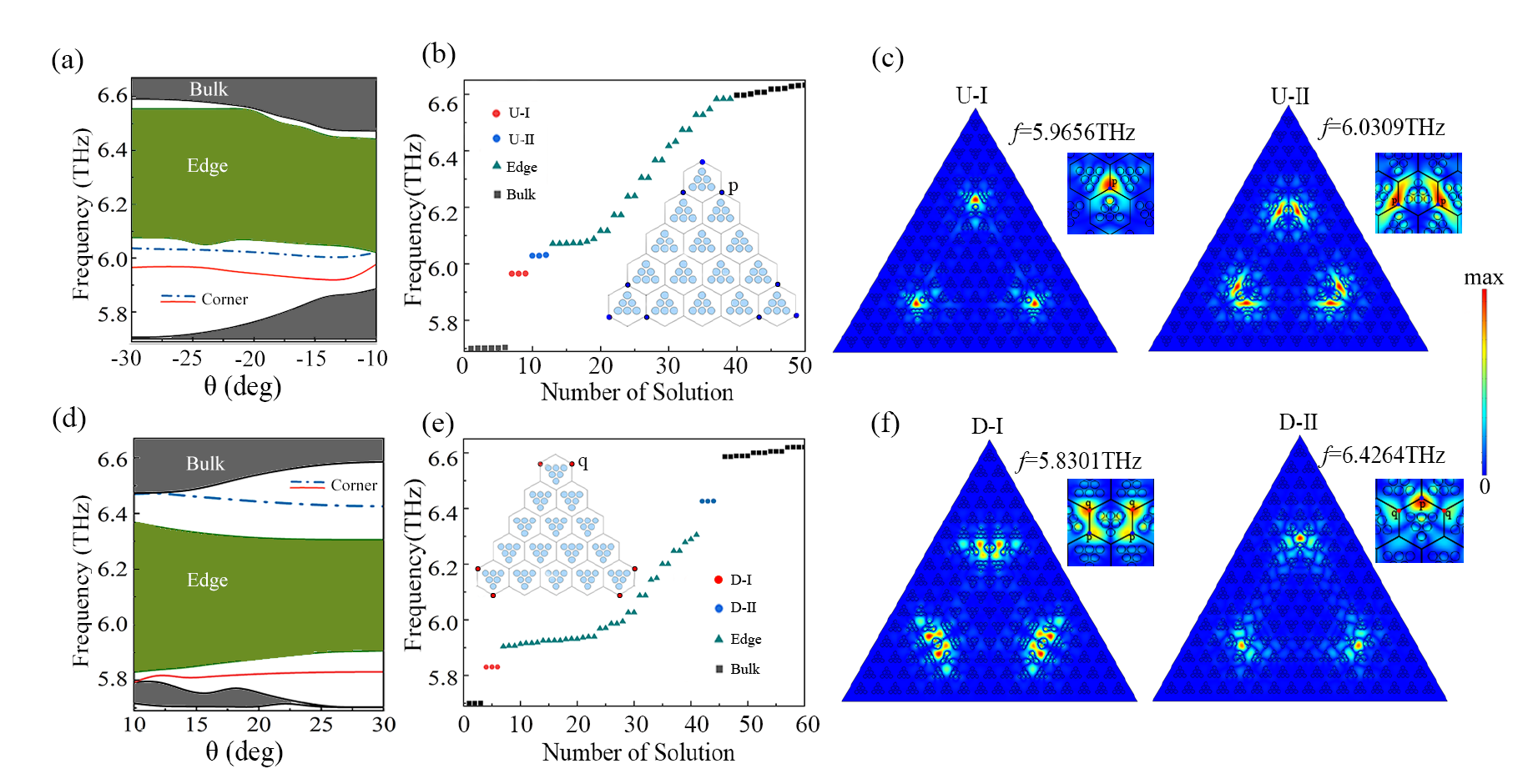}
\caption{\label{fig:wide}Up-corner states and down-corner states in a triangular nanodisk with opposite polarization. (a) Eigenfrequency evolution spectrum when $\theta \in\left(-30^{\circ}, -10^{\circ}\right)$. Red solid line for U-I corner states, and blue dot-dashed line for U-II corner states. (b) UPC eigenspectra of the bulk-edge-corner states, where the blue dots on the UPC indicate the positions of Wannier centers selected by $\mathrm{U}$-I and $\mathrm{U}$-II corner states. (c) Electric field distribution $\vert {E}_{z}\vert (x, y)$ of $\mathrm{U}$-I corner states at frequency $f=5.9656\mathrm{THz}$, and of $\mathrm{U}$-II corner state at frequency $f=6.0309\mathrm{THz}$. (d) Eigenfrequency evolution spectrums when $\theta \in\left(10^{\circ}, 30^{\circ}\right)$. Red solid line for D-I corner states, and blue dot-dashed line for D-II corner states. (e) DPC eigenfrequency distribution of the bulk-edge-corner states. The red dot of the DPC super unit model represents Wannier center selected by the down-corner states. (f) Electric field distribution of $\mathrm{D}$-I and $\mathrm{D}$-II corner states at frequencies $f=5.8301$ and $6.4264 \mathrm{THz}$, respectively.}
\end{figure*}  

Therefore our HOTI supports thus-defined corner states in the bandgap, which appear at the maximal Wyckoff positions of the unit cell. Generally in $C_{n}$-symmetric lattices, given a choice of unin cell, there exist special high-symmetry points in the unit cell, which are called maximal Wyckoff position(\emph{cf}. Sec.~$\mathrm{IV}$ of Supplemental Material~\cite{SM}). As in Eq.~\eqref{Pi}, the nontrivial second-order topology and emergence of the valley-selective corner states are theoretically characterized by the nontrivial bulk polarizations and the associated Wannier centers. Here the Wannier center refers to the center of the maximally localized Wannier function and for nontrivial polarization insulators, Wannier center is located at same position with the maximal Wyckoff position in the unit cell~\cite{2017d,2017Electric}.

\section{Numeric results and discussions}

To investigate the concept of valley-selective HOTI, we construct nanodisks made of two types of triangular lattices with distinct polarizations. When $\theta \in\left(-30^{\circ}, -10^{\circ}\right)$, the eigenspectra of our nanodisk is shown in Fig. 4(a). The two colored curves indicate the eigenfrequency functions with $\theta$ for the two types of vertices [Up-corner I ($\mathrm{U}$-I) and Up-corner II ($\mathrm{U}$-II) for shorthand respectively] 
Here, we refer to the PC with $\theta=-30^{\circ}$ as the up-triangular PC (UPC) and $\theta=30^{\circ}$ as the down-triangular PC (DPC). The eigenfrequencies of bulk-edge-corner in the UPC structure is shown in Fig.~4(b), where the $\mathrm{U}$-I and $\mathrm{U}$-II corner states both are triply degenerate. In insets of panels (b, e), Wannier centers are colored at the corners of the UPC structure. In our simulation, UPC is surrounded by DPC to interface an edge mode. As the electric field shows in Fig.~4(c), Wannier center representation, illustrated by the red dots $q$ in UPC structure, reveals the valley-selectivity of $\mathrm{U}$-I corner states. And the blue dots $q$ in the UPC structure reveals the valley-selectivity of the $\mathrm{U}$-II corner state. When $\theta \in\left(10^{\circ}, 30^{\circ}\right)$ the $\mathrm{D}$-I and $\mathrm{D}$-II corner states respectively appear below and above the edge state, as shown in Fig.~4(d). From the eigenfrequency distribution of the DPC structure, it is found that $\mathrm{D}$-I and $\mathrm{D}$-II corner states each also have three degenerate corner states, and the Wannier center configurations (in red dots) of the corner UPC structure as shown in Fig.~4(e). As the electric field in Fig.~4(f) shows, Wannier centers (\emph{cf}. p, q points in picture) of UPC and DPC are both fired by corner states. The reason is that when a DPC structure is surrounded by UPC, the corner states of the two models will be excited at the same time (\emph{cf}. Sec.~$\mathrm{V}$ of Supplementary Material~\cite{SM}). Moreover, the amplitude of $\mathrm{D}$-I corner electric field (f=5.8301THz) is higher than that of $\mathrm{D}$-II corner electric field (f=6.4264THz). We remark that the valley selectivity behaves globally, even though we only consider the UPC and the DPC here.
 
\begin{figure}
\includegraphics[width=0.5\textwidth]{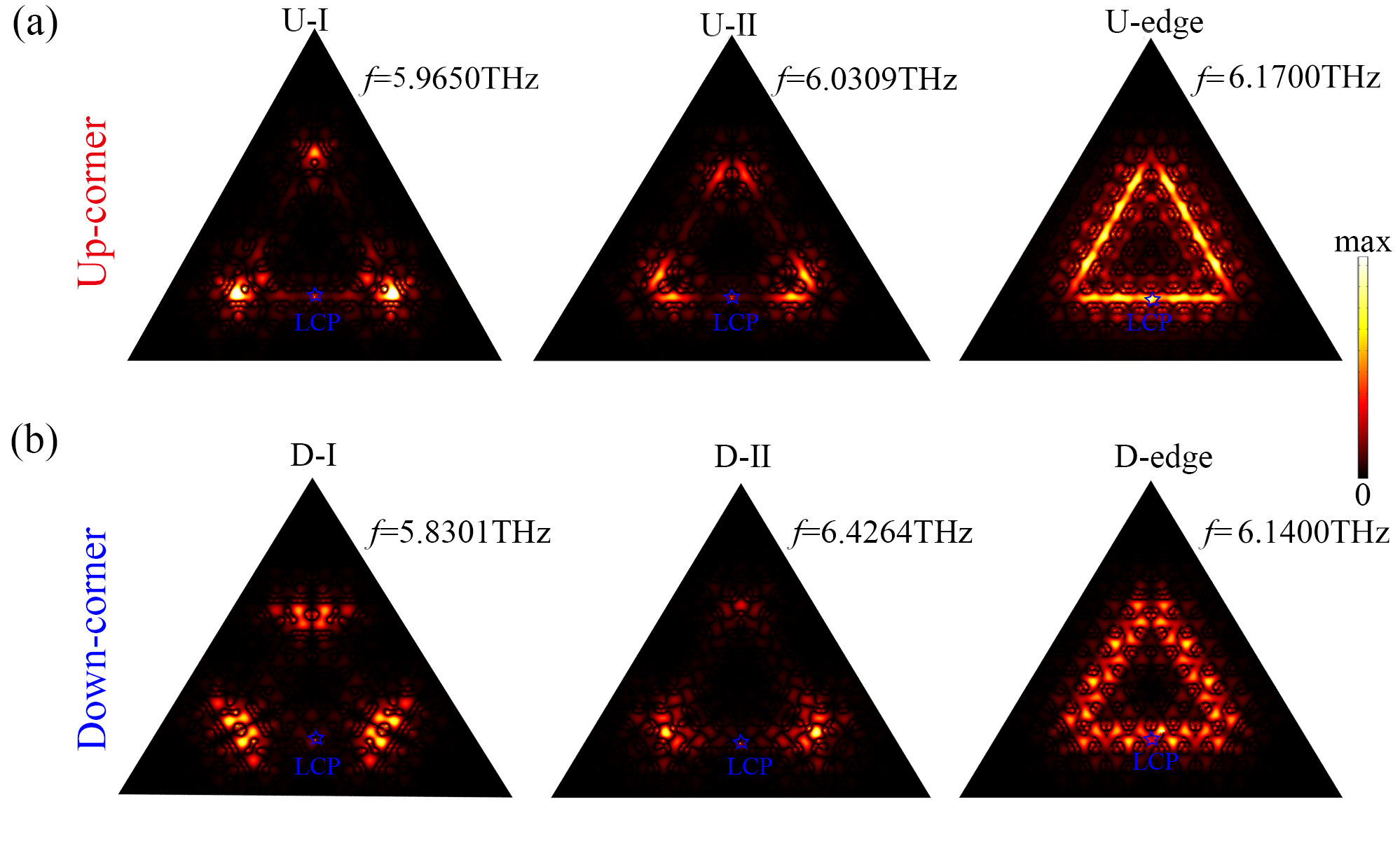}
\caption{\label{fig:epsart}
Simulated electric fields $\vert {E}_{z}\vert (x, y)$ for a configuration consisting of UPC [\emph{cf}.~inset of Fig.4(b)] and DPC [\emph{cf}.~inset of Fig.4(e)]. The purple pentagram at the bottom of configuration indicates a chiral OAM source. (a) Field distributions of the $\mathrm{U}$-I, $\mathrm{U}$-II corner and U-edge, where UPC is surrounded by DPC, are excited by LCP chiral sources with frequencies $f=5.9650, 6.0310, 6.1700{\rm THz}$. (b) Fields distributions of $\mathrm{D}$-I, $\mathrm{D}$-II corner and D-edge, where DPC is surrounded by UPC, are also excited by LCP sources with frequencies $f=5.8301, 6.4260, 6.1400{\rm THz}$.
}
\end{figure}

Now we set up full-wave simulation to verify the corner and edge states above in one, where the valley dependence of OAM chirality can exploited to achieve unidirectional excitation of valley chiral states. In Fig.~(5), we consider chiral line sources [in purple marker where we choose a left-handed circular polarized(LCP) OAM source] with a uniform magnitudes but chiral phase, which are fired near the bottom of our PhC with three zigzag boundaries. By switching the source frequency, we can directly control the appearance of edge and corner states as shown in Fig.~5(a-b). In the super unit where UPC is surrounded by DPC, $\mathrm{U}$-I and $\mathrm{U}$-II corner states are respectively excited at frequencies $f=5.9656 \mathrm{THz}$ and $f=6.0309 \mathrm{THz}$ at the same frequencies with Fig.~4(c). It is noted that corner states rely more sensitively on frequency parameters than edge ones do. Since the corner state transmits with loss, the electric amplitude of the corner state near the source remains higher than the further one. We choose frequency $f=6.1700 \mathrm{THz}$ to fire edge states, and our simulation shows that electromagnetic waves propagate smoothly along the interface even with sharp corners. It will promise new methods for streering electromagnetic waves along arbitrarily-cornered pathways~(\emph{cf}.~Sec.~$\mathrm{VII}$ of Supplementary Material~\cite{SM}). In the nanodisk where DPC is surrounded by UPC, $\mathrm{D}$-I and $\mathrm{D}$-II corner states are excited at $f=5.8301 \mathrm{THz}$ and $f=6.4264 \mathrm{THz}$, respectively. And the edge states are excited at $f=6.1400 \mathrm{THz}$. Our results then show that corner states can be selectively excited by tuning source frequency in addition to by valley selection. 

\section{\label{sec:level3}CONCLUSION}
In summary, we numerically realize a valley-type second-order topology due to unit-cell rotation characterized by the nontrivial bulk polarization. Specifically, the corner states are found to be valley dependent and therefore enable flexible control and manipulation on the wave localization. Thus topological switches by valley selection of corner states are numerically demonstrated in our paper. Our valley HOTI and the valley-selective corner states provide fundamental understanding on the interplay between higher-order topology and valley degree of freedom, which may find potential applications in valleytronics for future information carriers, such as waveguides, couplers and topological circuit switches in THz regime~\cite{2019Tunable,2019Mirror,2018VD,2019All,2017T,2020Terahertz,2021Demonstration, 2021All-dielectric}.


\begin{acknowledgments}

We are supported by Natural National Science Foundation (NSFC11804087, 12074108, 11704106, 12047501); Science and Technology Department of Hubei Province (2018CFB148); Hubei University (A201508, 030-090105); and Institute of Physics Carers’ Funds (IOP, U. K.). Our work is also partially supported by Hongque Innovation Center.

\end{acknowledgments}

\emph{Author contribution}. R. Z. and D.-H. X. proposed the idea. R. Z. performed the calculation, produced all the figures, and wrote the manuscript draft. H. L. and Y. L. lead the project and revised the whole manuscript thoroughly. R. Z., Y. L. and D.-H. X. put inputs together from all other coauthors in the manuscript revision. 

\emph{Disclosures}. The authors declare no conflicts of interest.



\providecommand{\noopsort}[1]{}\providecommand{\singleletter}[1]{#1}%

\end{document}